\documentclass[aps,prl,preprint,groupedaddress]{revtex4}

\usepackage{graphicx,amssymb,amsmath}
\usepackage[]{color}
\usepackage{ulem}

\begin{document}


\title{Experimental Realization of Type-II Dirac Fermions in PdTe$_2$ Superconductor}


\author{Han-Jin Noh}
\email{ffnhj@jnu.ac.kr}
\author{Jinwon Jeong}
\author{En-Jin Cho}
\affiliation{Department of Physics, Chonnam National University, Gwangju 61186, Korea}
\author{Kyoo Kim}
\affiliation{MPPC\_CPM, Pohang University of Science and Technology, Pohang 37673, Korea}
\author{B. I. Min}
\affiliation{Department of Physics, Pohang University of Science and Technology, Pohang 37673, Korea}
\author{Byeong-Gyu Park}
\affiliation{Pohang Accelerator Laboratory, Pohang University of Science and Technology, Pohang 37673, Korea}


\date{\today}

\begin{abstract}
A Dirac fermion in a topological Dirac semimetal is a quadruple-degenerate quasi-particle state with a relativistic linear dispersion.
Breaking either time-reversal or inversion symmetry turns this system into a Weyl semimetal that hosts double-degenerate Weyl fermion states with opposite chiralities.
These two kinds of quasi-particles, although described by a relativistic Dirac equation, do not necessarily obey Lorentz invariance, allowing the existence of so-called type-II fermions.
Recent theoretical discovery of type-II Weyl fermions evokes the prediction of type-II Dirac fermions in PtSe$_2$-type transition metal dichalcogenides, expecting an experimental confirmation.
Here, we report an experimental realization of type-II Dirac fermions in PdTe$_2$ by angle-resolved photoemission spectroscopy combined with {\it ab-initio} band calculations.
Our experimental finding makes the first example that has both superconductivity and type-II Dirac fermions, which turns the topological material research into a new phase.

\end{abstract}


\maketitle

The discovery of massless Dirac fermions in graphene/graphite\cite{Novoselenov, Zhou} and topological insulators (TIs)\cite{Kane1, HZhang} has ignited explosive researches on relativistic quasi-particles, topological properties of electronic structures, and their inter-connections in condensed matter physics\cite{KaneRMP}.
A crystalline system with time-reversal (TR) and inversion (I) symmetries, if a valence band and a conduction band touch each other at one point in the energy-momentum space by some reasons, can host a low energy quasi-particle excitation with linear dispersion, which is called a Dirac fermion\cite{Castro}.
Since the energy dispersion around the touching point (Dirac point) is linear with the crystal momentum, it is well described not by a non-relativistic Schr\"{o}dinger equation but by a relativistic Dirac equation for a massless particle with the Fermi velocity instead of the speed of light.
The stability of the Dirac point depends on the reason that makes the point touch.
If it is accidental, the point is fragile, but some kinds of symmetry in particular space groups are known to protect the Dirac point\cite{Young}.
In the case of symmetry protection, when TR- or I-symmetry is broken, the quadruple degenerate Dirac point is split into two doubly degenerate points (Weyl points) with opposite chiralities\cite{Wan}.

The generalized effective Hamiltonian for Dirac/Weyl fermions is given by $\hat{H}(\mathbf{k})=k_{i}v_{ij}\sigma_{j}$, where $i=(x,y,z)$, $j=(0,x,y,z)$, $k_i$'s are the wave vectors, and $\sigma_{j}$'s the identity/Pauli matrices.
The coefficients $v_{ij}$'s ($j\neq0$) are proportional to the velocity of the quasi-particle in each momentum/spin direction, and $v_{i0}$'s to the energy of the Dirac/Weyl point.
The energy dispersion relation of the Hamiltonian is given by the two solutions of the quadratic eigenvalue equation for the 2$\times$2 matrix, $\varepsilon_{\pm}(\mathbf{k})$=$k_{i}v_{i0} \pm \sqrt{\sum_{j\neq0} (k_{i}v_{ij})^2 }$=$T(\mathbf{k})\pm U(\mathbf{k})$.
If $|T(\mathbf{k})| > |U(\mathbf{k})|$ for a $\mathbf{k}$ direction, the Dirac cone is so tilted that the electron and hole cones touching each other are cut by the constant Fermi energy plane, making the Dirac/Weyl points to cross Fermi lines instead of Fermi points.
It actually corresponds to a kind of structured Weyl points proposed by Xu {\it et al.}\cite{Xu}.
In contrast to the type-I case of $|T(\mathbf{k})| < |U(\mathbf{k})|$, this type-II case violates the Lorentz invariance and has different topology\cite{Soluyanov}.
The concept of type-II Weyl fermion proposed by Soluyanov {\it et al.} immediately extended to Dirac fermion by Huang {\it et al.}\cite{Huang}, predicting the existence of type-II Dirac fermions in PtSe$_2$ class systems as spin-degenerate counterparts of type-II Weyl fermions, and an experimental test on PtTe$_2$ was reported\cite{Yan}.

In this Letter, by combining angle-resolved photoemission spectroscopy (ARPES) and {\it ab-initio} band calculations we report the experimental discovery of type-II Dirac fermions in PdTe$_2$, which establishes the first example that has both superconductivity and type-II Dirac fermions.
Our photon-energy dependent and circular dichroic (CD) ARPES measurements clearly reveal that this system has type-II Dirac points with a binding energy of $\varepsilon_{D-II}$=0.5 eV at $\mathbf{k}$=(0,0,$\pm0.4$) in reciprocal lattice unit.
Also, this system has another kind of Dirac point with a binding energy of $\varepsilon_{SD}$=1.7 eV in a (001) surface, originating from a non-trivial Z$_2$ invariant.
All of our experimental findings are supported by the band calculation based on a density functional theory (DFT) including the spin-orbit coupling.

Layered transition metal dichalcogenide (TMD) is one of the most important material groups that shows plenty physical phenomena such as charge density waves\cite{Wilson, Valla}, superconductivity\cite{Revolinsky1, Revolinsky2}, metal-insulator-transition\cite{Kunes}, unsaturate magneto-resistance\cite{Ali}, etc.
One of the TMDs, PdTe$_2$ is an intermetallic compound and is known to become a superconductor below T$_c$=1.7 K\cite{Jellinek}, which is comparable to other TMD superconductors.
A fully relativistic band calculation with spin-orbit splitting was performed by Jan and Skriver in 1970's\cite{Jan}, which successfully explained most of the de Haas-van Alphen results\cite{Dunsworth}.
The first experimental observations of the electronic structure of PdTe$_2$ by photoemission and scanning tunneling microscopy were reported to show that the Te atoms are the main current source for low bias voltages\cite{Orders, Ryan}.
Recently, high resolution ARPES studies combined with DFT calculations were reported by Liu {\it et al.}, where the electronic structure was measured in detail using a He discharge lamp and a topologically non-trivial surface state was observed at the $\Gamma$ point in a surface Brillouin zone (BZ)\cite{Liu1, Liu2}.
However, due to the limited photon energies and non-polarity, critical experimental evidence for topological properties such as $k_z$ band dispersions and spin texture is lacking.
In the following, we first provide the detailed electronic structure measured by photon energy dependent ARPES and CD-ARPES together with the calculated band structure, then we present undisputable experimental evidence for the existence of type-II Dirac fermions in PdTe$_2$.


\begin{figure}
\includegraphics[width=16.0 cm]{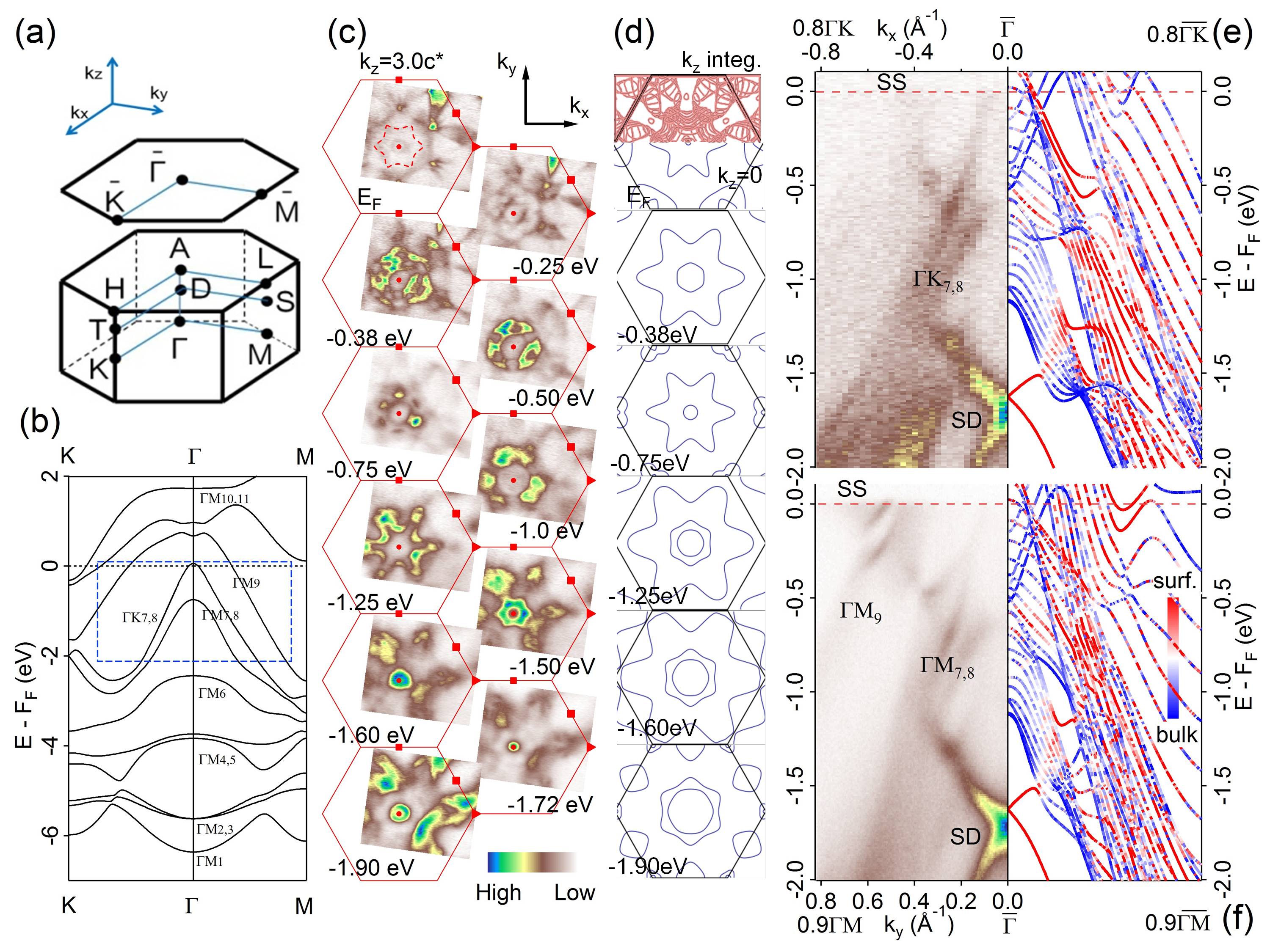}
\caption{\label{fig1}
\textbf{ARPES-measured electronic structure of PdTe$_2$.}
(a) Bulk and (001) surface Brillouin zones for a hcp structure.
(b) Calculated bulk band structure along the K-$\Gamma$-M symmetric line by the DFT method. The blue dotted box is the region of ARPES measurement.
(c) Constant energy maps measured at energies ranging from 0 ($E_F$) to -1.90 eV using $\hbar\omega$=36.0 eV, which corresponds to $k_z$=3.0c* (c*=2$\pi$/$c$). The red hexagons are BZs.
(d) Corresponding isoenergy contours by DFT calculations. The upper half of the topmost contour is a $k_z$-integrated Fermi surface map from -0.5c* to 0.5c*, and the others are for $k_z$=0.
(e,f) ARPES image (left) and calculated surface band structure (right) in the $\overline{\Gamma}$-$\overline{\textrm{K}}$/$\overline{\Gamma}$-$\overline{\textrm{M}}$ direction. The surface states (SS) and the surface Dirac cone (SD) show strong surface character (red color) in the calculations.
}
\end{figure}

The crystal structure of PdTe$_2$ is of CdI$_2$ type ($P\overline{3}m1$) with a unit cell consisting of one Pd atom and two Te atoms.
The corresponding bulk BZ and (001) surface BZ are shown in Fig.~\ref{fig1}(a).
The lattice constants are $a$=4.0365 and $c$=5.1262 \AA, so the ratio of $c$ to $a$ is $c/a$=1.27, which is far below the value ($\sqrt{8/3}$=1.63) of an ideal $hcp$ structure, implying relatively strong hybridization along the $c$-direction\cite{Finlayson}.
The calculated bulk band structure based on DFT along the K-$\Gamma$-M symmetric line is displayed in Fig.~\ref{fig1}(b).
The constant energy maps for $k_z$=0 measured at energies from zero (E$_F$) to -1.9 eV on a (001) cleaved surface and the corresponding isoenergy contours by DFT are shown in hexagonal BZs in Fig.~\ref{fig1}(c) and \ref{fig1}(d), respectively.
The upper half of the topmost contour drawn with brown lines in Figs.~\ref{fig1}(d) is a theoretical Fermi surface integrated over $k_z$ from -0.5c* to 0.5c* (c*=$2\pi/c$).
At E=E$_F$, a small circle, a faint small hexagon (red dotted), a large hexagon around $\Gamma$, and a petal-shaped structure around K are seen in the ARPES data.
Meanwhile, there are a small circle, an asterisk around $\Gamma$, and two rounded triangles around K in the corresponding contour.
If we lower a measuring energy, the small circle and the hexagon get larger, but the petals get shrunk, which implies that the structures around $\Gamma$ are hole pockets and that the structures around K are electron pockets.
They are also checked directly in the energy dispersive ARPES images for $\overline{\Gamma}$-$\overline{\textrm{K}}$ and $\overline{\Gamma}$-$\overline{\textrm{M}}$ directions as shown in Figs.~\ref{fig1}(e) and \ref{fig1}(f), respectively.

These features are consistent with the DFT calculations in some degrees, but a noticeable difference between the experimental and theoretical results is that the former has more abundant structures than the latter.
This looks clear when we compare the dispersive ARPES images with the bulk band calculations along the K-$\Gamma$-M line in Fig.~\ref{fig1}(b).
The blue dotted box in Fig.~\ref{fig1}(b) indicates the ARPES-measured region, where there are only three hole-like bands ($\Gamma$M$_7$/$\Gamma$K$_7$, $\Gamma$M$_8$/$\Gamma$K$_8$, and $\Gamma$M$_9$).
Meanwhile, a lot of extra structures appear in the ARPES images both in $\Gamma$-K and $\Gamma$-M directions.
We suspect that these extra structures consist of surface states and $k_z$-projected bulk states in ARPES.
The surface state contribution can be checked by comparing the energy-dispersive ARPES data with the surface band calculations based on a slab model as shown in the right side of Figs.~\ref{fig1}(e) and \ref{fig1}(f).
The structures denoted as $SS$ and $SD$ in the ARPES data, are well reproduced in the calculations with strong surface character (red color), while $\Gamma$M$_7$/$\Gamma$K$_7$, $\Gamma$M$_8$/$\Gamma$K$_8$, and $\Gamma$M$_9$ are consistent with the bulk states (blue color).
The $k_z$ projection effect, which is an intrinsic shortcoming of ARPES for a 3-dimensional system study\cite{Strocov}, can be seen in the integrated Fermi surface map over $k_z$ as shown in the topmost contour of Fig.~\ref{fig1}(d).
The petal-shaped structures in the ARPES image are well reproduced in the simulation.
This indicates that the system has non-negligible band dispersion in the $k_z$ direction, being consistent with the fact that PdTe$_2$ has the $hcp$ crystal structure with a small $c/a$ ratio.
Actually, the projection effect is so conspicuous in PdTe$_2$ that careful data interpretation is essential to the investigation of the electronic structure in $k_z$ direction.

\begin{figure}
\includegraphics[width=14.0 cm]{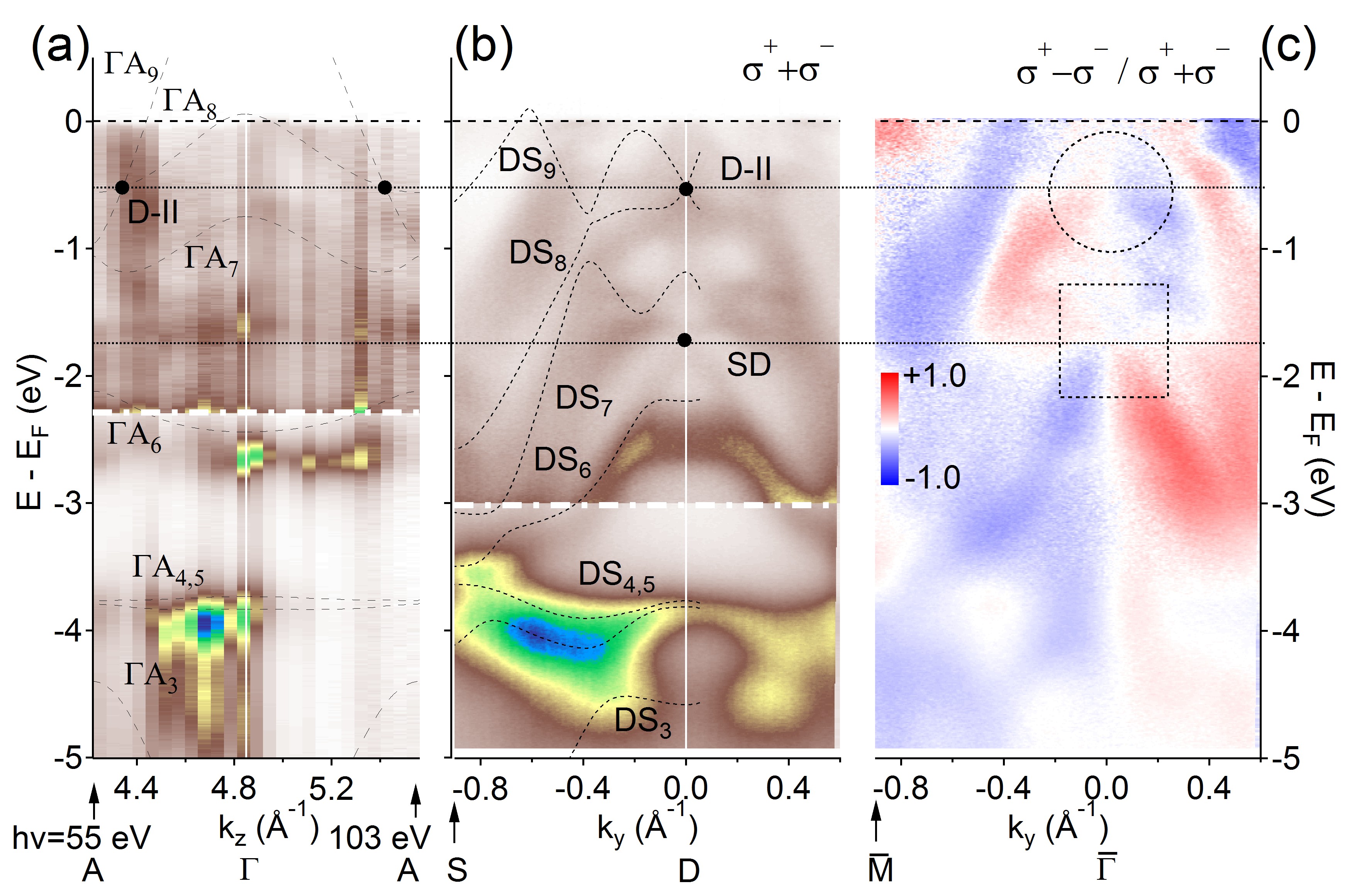}
\caption{\label{fig2}
\textbf{Photon energy dependence and circular dichroism of the ARPES data.}
(a) ARPES image of PdTe$_2$ along the A-$\Gamma$-A line measured with varying photon energy from 55 to 103 eV.
The thin dotted curves are calculated bulk bands along the same line.
(b) ARPES image along the D-S line (k$_z$=3.6c*) obtained by summing two ARPES images using circular light with opposite polarization.
The dotted curves are calculated bulk bands along the same line.
(c) Normalized circular dichroic ARPES image along the D-S line.
A noticeable difference is observed in dichroic reversal between the D-II region (dotted circle) and the SD region (dotted square).}
\end{figure}

Among the extra-structures in the ARPES data, first attention is paid to the $SD$ state.
It looks like a typical Dirac cone in a TI, and the theoretical study\cite{Huang} also predicted that PdTe$_2$ has nontrivial $Z_2$ topology as well as that the type-II Dirac point exists at $\mathbf{k}_{D-II}$=(0,0,0.406c*) with the binding energy $\varepsilon_{D-II}=0.545$ eV.
In order to experimentally confirm the topological character of $SD$, and to find the Type-II Dirac point, photon energy dependent ARPES and CD-ARPES measurements are exploited as shown in Fig.~\ref{fig2}.
The band dispersion along the A-$\Gamma$-A line in Fig.~\ref{fig2}(a) is measured by tuning the photon energy from 55 to 103 eV by every other electron-volt.
For comparison, the bulk band calculations are shown with the thin dotted lines.
Even though the $k_z$ momentum resolution is not so high due to the $k_z$ projection effect and the matrix-element effect in photoemission, the bulk bands $\Gamma$A$_3$ - $\Gamma$A$_9$ are captured in the ARPES data.
Also, the non-dispersive $SD$ state at $\sim$1.7 eV below $E_F$ is clearly observed, which strongly indicates that the state are confined to a (001) surface region.

The $k_z$ scan suggests that the type-II Dirac point can be caught when $\hbar\omega$=61 eV photons are used as shown in Fig.~\ref{fig2}(b).
The corresponding band calculations are drawn with dotted lines.
In our calculation, the bulk Dirac point is predicted to exist at (0,0,0.401c*), showing a negligible difference from Ref.\cite{Huang}.
When we compare the ARPES image with the calculations, the Dirac point (D-II) at $\varepsilon$=-0.5 eV is discernable, but other structures with surface origin or with different $k_z$ values are also prominent.
Among them, the surface states can be experimentally identified by CD-ARPES because an inversion symmetry breaking at the surface induces spin splitting of the surface states.
Even though an absolute spin structure cannot be obtained from CD-ARPES due to dominant final state effects, a relative spin structure, or a projected spin structure to a plane in momentum space can be obtained\cite{Scholz, Arrala}.
Our normalized CD-ARPES result is displayed in Fig.~\ref{fig2}(c), where spin-split surface states get dominant.
Here, the normalized CD is obtained by the difference divided by the sum in the ARPES intensities measured with right ($\sigma^{+}$) and left ($\sigma^{-}$) circular polarized light, respectively; that is CD = ($\sigma^{+}-\sigma^{-}$)/($\sigma^{+}+\sigma^{-}$).
The actual CD values in the ARPES data range from -0.6 to +0.6.

A contrasting feature is observed in SD and D-II states.
In the SD state denoted with a dotted box in Fig.~\ref{fig2}(c), the dichroic signal is reversed with respect to the SD point just as a typical spin-momentum locked topological surface state Dirac cone.
In contrast, the dichroic signal of D-II is reversed not with respect to the D-II point (dotted circle) but with respect to the vertical $k_y$=0 line.
Also, it almost disappears at the upper part of the D-II state while showing considerable intensities in the lower part.
The weak dichroic signal in the upper cone indicates that D-II is a quadruple-degenerate bulk Dirac point.
If TR- or I-symmetry were broken, the D-II point would be split into two Weyl points with opposite chiralities and a strong dichroic signal would appear as in the SD region.
Meanwhile, the relatively strong dichroic signal in the lower part of D-II is the result of a strong mixing between the topological surface state and the $DS_8$ bulk band.
This can be directly checked if we compare the dichroic image with the surface band calcuations in Fig.~\ref{fig1}(f).
The high surface character regions coincide with the high dichroism regions.
In this interpretation, it is worth mentioning that the contrasting dichroism reversal behavior is a result from neither a geometric effect nor a final state effect.
A geometric effect in CD-ARPES cannot selectively reverse the dichroism in one common experimental setup.
Exclusion of final state effects, in principle, requires exact calculations of the matrix elements in photoemission process\cite{Arrala}, but in short, a selective dichroism reversal in a small energy range ($\sim$1 eV) cannot occur at continuum-like final states well above the Fermi level ($\sim$50 eV).

\begin{figure}
\includegraphics[width=16.0 cm]{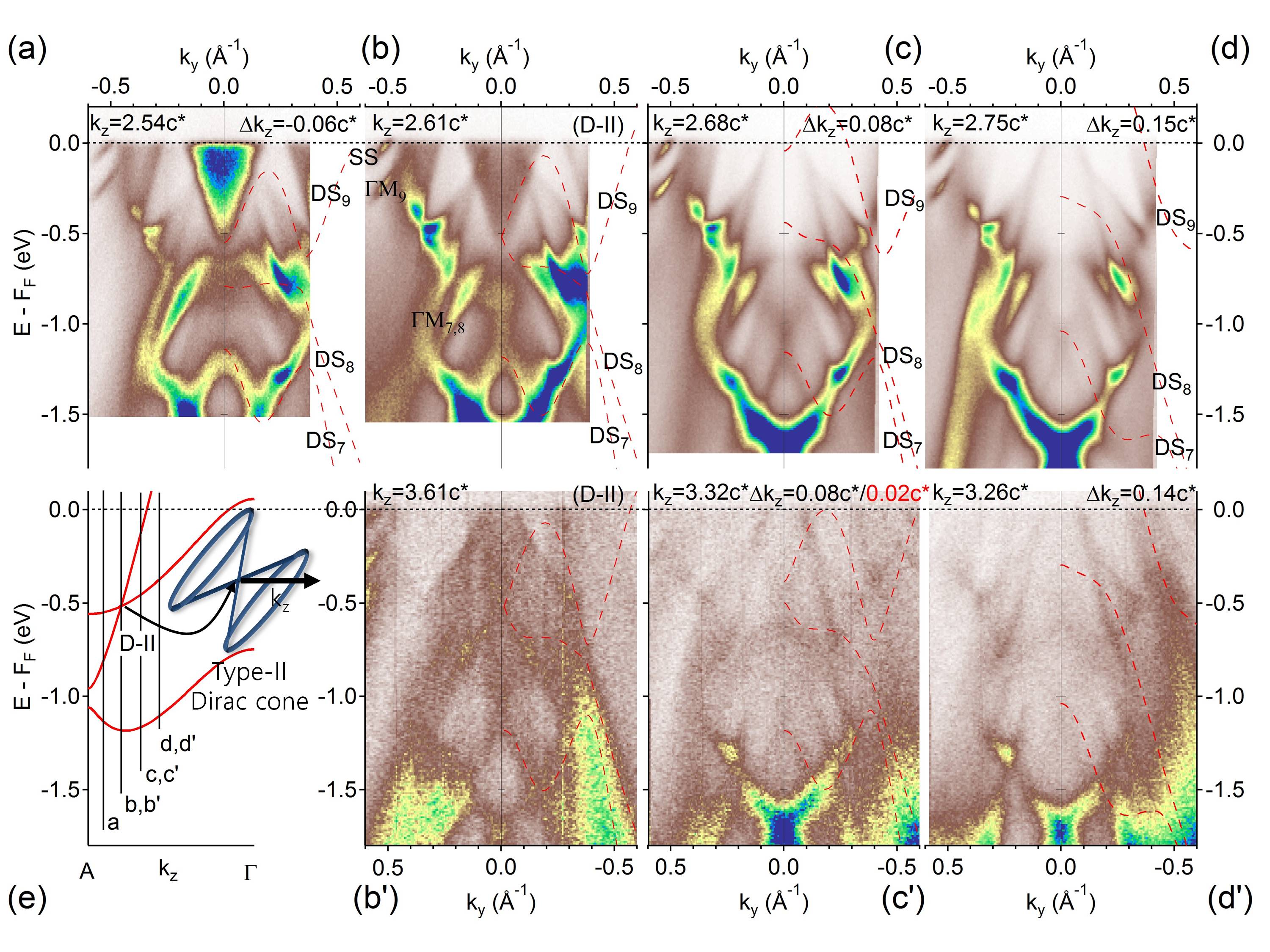}
\caption{\label{fig3}
\textbf{Type-II Dirac fermion in PdTe$_2$ evidenced by ARPES.}
(a,b,c,d) ARPES images along the k$_y$ direction at k$_z$=2.54c* ($\Delta{k}$=-0.06c*), 2.61c* (D-II), 2.68c* ($\Delta{k}$=0.08c*), and 2.75c* ($\Delta{k}$=0.15c*). The red dashed lines are the calculated bands at the corresponding k$_z$ value.
(b',c',d') ARPES images measured in the next BZ at k$_z$=3.61c* (D-II), 3.32c* ($\Delta{k}$=0.08c*), 3.26c* ($\Delta{k}$=0.14c*).
In (c'), the calculated bands are for $\Delta{k}$=0.02c*.
(e) Calculated bands along the A-$\Gamma$ line and schematic type-II Dirac cone. The vertical lines denote the measuring k$_z$ point of the ARPES images in (a-d').
}
\end{figure}

In order to experimentally confirm the type-II conditions of the D-II point, we obtained high resolution ARPES images with $\sim$2 eV energy window by various important photon energies as shown in Figs.~\ref{fig3}(a)-(d').
Figure~\ref{fig3}(e) schematically shows the relative $k_z$ positions of the ARPES data in a reduced zone scheme.
Also, see Fig.~\ref{fig1}(a) for the $\mathbf{k}$-point notation.
In these ARPES data, the strong spectral weight of the surface states and the $k_z$-projected bulk bands make it hard to follow the $k_z$ dispersion of the DS$_8$ and DS$_9$ bands, but we already identified the surface state SS and three bulk bands $\Gamma$M$_{7,8,9}$ in Fig.~\ref{fig1}(f), which help us to assign DS$_8$ and DS$_9$.
At $k_z$=2.54c* ($\Delta{k}\equiv$$k_{D-II}-k_z$=-0.06c*), DS$_8$ is at about -0.7 eV, showing a clear gap between DS$_{8,9}$.
The DS$_9$ band, which is an electron pocket-like structure at $k_y$=0, shows a little deviation from the calculations around $k_y$=0.2 {\AA}$^{-1}$, but the minimum point is almost the same as the theoretical value.
At $k_z$=2.61c* ($\cong k_{DII}$), the two bands almost contact each other at the $k_y$=0 point.
The contact point and the Dirac cone are more prominent in $k_z$=3.61c* data ($\cong k_{DII}$ in next BZ) as shown in Fig.~\ref{fig3}(b').
Unfortunately, from this value of $k_z$, the spectral weight of DS$_8$ and DS$_9$ near $k_y$=0 becomes extremely low that we cannot keep track of DS$_8$ and DS$_9$ in Figs.~\ref{fig3}(c,d), so we employed higher photon energies to reach the next BZ as shown in Figs.~\ref{fig3}(c',d').
When $\Delta k$=0.08c* ($k_z$=3.32c*) in Fig.~\ref{fig3}(c'), the Dirac point splits by $\sim$0.1 eV and the both points move upward.
In this figure, the calculated bands are for $\Delta k$=0.02c*, while the photon energy corresponds to $\Delta k$=0.08c*.
When $\Delta k$=0.16c* ($k_z$=2.76c* or 3.26c*) in Figs.~\ref{fig3}(d,d'), DS$_9$ disappears above the Fermi level, and the top position of DS$_8$ is around -0.3 eV.
Even though the dispersion of DS$_8$ along $k_y$ direction shows considerable deviations from that of calculations, the same upward dispersion of both DS$_8$ and DS$_9$ along $k_z$ direction shows that the inequality $|T(k)| > |U(k)|$ holds, satisfying the type-II condition of the D-II Dirac point.

Up to now, by ARPES combined with first principle calculations, we provide experimental evidence that the electronic structure of PdTe$_2$ can host type-II Dirac fermionic quasiparticles at $\mathbf{k}$=(0,0,$\pm$0.4c*) with $\varepsilon_{D-II}$=-0.5 eV.
This actually establishes that PdTe$_2$ is the first example that has both superconductivity and type-II Dirac fermions, providing a possible platform of research for interactions between superconducting quasiparticles and type-II Dirac fermions.
Straightforward tactics to make the interactions measurable is to lower the Fermi level to near D-II point without losing the superconductivity, and there is a good feature to this direction, i.e. a sharp structure in the density of states near D-II point.
Thus, a successful Fermi level tuning may lead to another prototypical system like Cu$_x$Bi$_2$Se$_3$ in research of topological superconductors\cite{Hor}.
Also, the relatively small binding energy of the D-II point makes it feasible to observe the anticipating peculiar magneto-transport properties in its present form\cite{Soluyanov,Zyuzin,Fei}.
Since this quasiparticle has no counterpart in particle physics owing to the violation of the Lorentz invariance, a finding of unconventional property may step into new physics.
Hopefully, our experimental realization of type-II Dirac fermions in PdTe$_2$ can turn the research of Dirac semimetals into a new phase.

\subsection{Experimental methods}

The PdTe$_2$ single crystals were grown by a Te self-flux method.
A mixture of Pd (99.95$\%$) and Te (99.99$\%$) powders with a 1:10 ratio was loaded in an evacuated quartz ampoule.
The ampoule was heated up to 890$^{\circ}$C for 30 hours, followed by slow cooling to 630 $^{\circ}$C at a rate of 4 $^{\circ}$C/h, then furnace-cooled to room temperature.
The crystals were isolated from the Te flux by heating the ampoule to 480 $^{\circ}$C to melt down the flux.
The single phase of the crystals was checked by X-ray diffraction, and the resistivity was measured at a temperature range of 2 - 300 K to confirm the metallic transport property.

The ARPES experiments were performed at the 4A1 beamline of the Pohang Light Source with a Scienta R4000 electron spectrometer and $\hbar\omega$=22$\sim$110 eV photons\cite{HDKim}.
The crystals were cleaved {\it in situ} by a top-post method at 60~K under $\sim7.0\times 10^{-11}$ Torr.
The total energy (momentum) resolution of ARPES data is $\sim$20~meV ($\sim$0.01~\AA$^{-1}$).
For $k_z$ mapping, we used the relation $k_z=\hbar^{-1}\sqrt{2m(E_{kin}\cos^2{\theta}+V_0)}$, where $m$, $E_{kin}$, $\theta$, and $V_0$ are the electron mass, the kinetic energy of photoelectrons, the emission angle from a sample normal, and the inner potential, respectively\cite{Damascelli_ARPES}.
For PdTe$_2$, the inner potential was estimated to be 19.0$\pm$0.3 eV from our photon energy dependent ARPES study.
This gives $\Delta{k_z}$$\sim$0.02c*.
For CD-ARPES, all experimental geometry was fixed, and the photon polarization was reversed using the elliptically polarizing undulators in the beamline.

For the bulk band structure calculations, we have used the full-potential linearized augmented plane wave band method with local orbitals, implemented in the DFT package, Wien2k\cite{Wien2k}.
The relativistic spin-orbit interaction of heavy atoms Pd and Te were treated by a second variational method in a relatively large window of 5 Rydbergs.
24$\times$24$\times$16 $\mathbf{k}$-points in full BZ are used for the reciprocal space integration.
For the surface band structure calculation, we have constructed a slab consisting of 11 layers of PdTe$_2$ separated by a vacuum regions of about 30 \AA~ thickness, then we relaxed the structure, utilizing pseudopotential planewave DFT package, VASP\cite{VASP1,VASP2}.
We kept in-plane dimensions of slab unit cell for the structure relaxation.
Surface band dispersions are calculated using Wien2k with relaxed structures.
Spin-orbit interactions are included, and 24$\times$24$\times$1 $\mathbf{k}$-points in the full surface BZ are used for all the surface calculations.


\subsection{Acknowledgments}
This work was supported by the National Research Foundation (NRF) of Korea Grant funded by the Korean Government (MEST) (Nos. 2010-0010771, 2013R1A1A2058195, 2016R1D1A3B03934980, 2016R1D1A1B02008461, and 2016K1A4A4A01922028).
The experiments at PLS and the theoretical calculations were supported in part by POSTECH and by the KISTI superconducting center (No. KSC-2015-C3-062).
Part of this work were performed using facilities at IBS Center for Correlated Electron Systems, Seoul National University.



\end{document}